\documentclass[runningheads]{llncs}

\makeatletter
\newcommand{\@chapapp}{\relax}%
\makeatother

\usepackage[title,toc,titletoc,header]{appendix}

\usepackage[utf8]{inputenc}
\usepackage{amssymb}
\usepackage{amsmath}
\usepackage{algorithm}
\usepackage{algorithmic}
\usepackage{graphicx}
\usepackage{hyperref}
\usepackage{listings}
\usepackage{subfig}
\usepackage{booktabs}
\usepackage{multirow}
\usepackage{colortbl}
\usepackage{marvosym}
\usepackage[moderate]{savetrees}

\begin{document}
\title{Identifying Near-Optimal Single-Shot Attacks on ICSs with Limited Process Knowledge}
\titlerunning{Identifying Attacks on ICSs with Limited Knowledge}
\author{Herson Esquivel-Vargas\inst{1}\textsuperscript{(\Letter)} \and
  John Henry Castellanos\inst{2} \and \\
  Marco Caselli\inst{3} \and
  Nils Ole Tippenhauer\inst{4} \and
  Andreas Peter\inst{1,5}
}
\institute{
  University of Twente, Enschede, The Netherlands\\
  \email{\{h.esquivelvargas,a.peter\}@utwente.nl}
  \and
  Singapore University of Technology and Design, Singapore\\
  \email{john\_castellanos@mymail.sutd.edu.sg}
  \and
  Siemens AG, Munich, Germany\\
  \email{marco.caselli@siemens.com}
  \and
  CISPA Helmholtz Center for Information Security, Saarbr\"{u}cken, Germany\\
  \email{tippenhauer@cispa.de}
  \and
  University of Oldenburg, Oldenburg, Germany
}
\authorrunning{H. Esquivel-Vargas et al.}
\maketitle
\vspace{-0.5cm}
\begin{abstract}
Industrial Control Systems (ICSs) rely on insecure protocols and devices to monitor and operate critical infrastructure. Prior work has demonstrated that powerful attackers with detailed system knowledge can manipulate exchanged sensor data to deteriorate performance of the process, even leading to full shutdowns of plants. Identifying those attacks requires iterating over all possible sensor values, and running detailed system simulation or analysis to identify optimal attacks. That setup allows adversaries to identify attacks that are most impactful when applied on the system for the first time, before the system operators become aware of the manipulations. 

In this work, we investigate if constrained attackers without detailed system knowledge and simulators can identify comparable attacks. In particular, the attacker only requires abstract knowledge on general information flow in the plant, instead of precise algorithms, operating parameters, process models, or simulators. We propose an approach that allows single-shot attacks, i.e., near-optimal attacks that are reliably shutting down a system on the first try. The approach is applied and validated on two use cases, and demonstrated to achieve comparable results to prior work, which relied on detailed system information and simulations.
\end{abstract}

\section{Introduction}
Attacks on Industrial Control Systems (ICSs) have been thoroughly investigated in the post-Stuxnet era.
Different initiatives such as Mitre's \emph{ATT\&CK for ICS} and SANS Institute's \emph{ICS Cyber Kill Chain} have emerged to systematically analyze ICS attacks~\cite{assante2015industrial,attack_ics}.
After a decade of research, it became clear that the difficulty of ICS attacks is not the execution itself but the preparation~\cite{DBLP:conf/ccs/GreenKA17}.
The latter stages of an ICS attack are considered easy because most industrial protocols lack security services such as message encryption and authentication, Programmable Logic Controllers (PLCs) remain operational for years without software updates, unprotected Human-Machine Interfaces (HMIs), and many more.
The difficulty about preparing an ICS attack is due the complexity of ICSs that can have hundreds of components (e.g., sensors, setpoints, actuators) and knowing \emph{what to target} and \emph{how to target it} is considered far from trivial.

Most research on ICSs security argues, or simply assumes, that detailed knowledge about the system is a requirement for a successful compromise.
In this work, we challenge such a claim by considering an attacker whose goal is to destabilize the physical process using only limited process knowledge.
Despite this preparation constraint, the attack (1) must target one component only once (single-shot); and (2) must have a fast effect after the attack execution (near-optimal).
The first requirement aims at executing stealthy and hard to detect attacks.
The second requirement discards attacks whose impact take a long time, increasing the chance of detection and defensive reaction.
More precisely, we call `near-optimal' the top-3 fastest attacks under the same system and attack conditions.

Previous research has identified the data sources required by an attacker to prepare a successful attack against ICSs~\cite{DBLP:conf/ccs/GreenKA17}.
For instance, PLC configuration, HMI/Workstation configuration, historian configuration, network traffic,  system/component constraints, and piping and instrumentation diagrams (P\&IDs).
The main argument being that only through the combination of multiple data sources, an attacker is able to reach the level of ``process comprehension'' required to launch an attack.
Previous works confirm this claim.
We analyzed the attack preparation phase of several papers and categorized their requirements according to the data sources presented in~\cite{DBLP:conf/ccs/GreenKA17}.
Table~\ref{tab:AttackComparison} shows that previous works use multiple data sources to prepare an ICS attack.

In this work, we propose a new approach that only requires limited knowledge of the system's architecture, and still allows to identify \emph{near-optimal single-shot} attacks.
Our approach requires an abstract representation of the information flow in the system - which sensors and setpoints influence which control decision, which actuators are controlled by which control function.
We show how to obtain this  information (e.g., from P\&IDs), and how to express it in an abstract graph representation.
We  then leverage the accumulated knowledge on IT software weaknesses and apply it to the ICSs domain.
In particular, we use weaknesses from Mitre's Common Weakness Enumeration (CWE) database,
and translate them to specific graph patterns.
Our goal is to look for these patterns in the ICS graph to identify suitable targets.
From this set of targets, the attacker picks one (e.g., randomly) and executes his `one-shot' attack.

We evaluate our approach using a simulated ICS known as the Tennessee Eastman Plant (TEP).
We used two different implementations of the TEP and build the corresponding CCL graphs.
Our analysis of the graphs reveals weaknesses which we hypothesize are the components in best capacity to compromise the availability of the plant.
We then execute simulations to test our hypotheses.
We found out that there is a correlation between the targets automatically chosen by our approach and the components that, according to the simulations, are prone to cause a shutdown of the plant.
Although different security aspects of the TEP have been studied in the past,
most of them have run simulations under very limited conditions~\cite{DBLP:journals/ijcip/HuangCALTS09,DBLP:conf/nordsec/KrotofilC13}.
To gain more confidence in our results, we executed 748 individual simulations under 14 different conditions that account for 2.52 years of simulated time.

Summarizing, our two main contributions are as follows.
(1) We present a novel method to identify near-optimal single-shot attacks on ICSs.
Unlike previous offensive approaches, our work uses limited process knowledge, thus, offering new insights about the actual security risks in ICSs.
(2) To validate our approach, we executed, documented, and published the most extensive set of simulated attacks on the TEP to date.
The code, results of each simulation, and screenshots are provided in the corresponding repositories.

\begin{table}[tb]
  \centering
  \small
  \caption{Comparison of the required attacker knowledge in previous works. Data sources are based on the process knowledge data source taxonomy from~\cite{DBLP:conf/ccs/GreenKA17}. P\&ID: Piping and Instrumentation Diagram, PLC: Programmable Logic Controller, HMI: Human-Machine Interface, WS: Workstation.}
  \label{tab:AttackComparison}
\setlength\tabcolsep{2pt} 
  \begin{tabular}{@{}lccccccccccc@{}}  
       \toprule
       \multirow{2}{*}{\textbf{Data source}} & \multicolumn{11}{c}{\textbf{Work}} \\
       \cmidrule{2-12}
       & \cite{adepu2016distributed} & \cite{ahmed2018noise}  & \cite{DBLP:conf/ccs/CardenasALHHS11} & \cite{chen2019learning} & \cite{chen2020active} & \cite{DBLP:journals/ijcip/HuangCALTS09} &\cite{DBLP:conf/nordsec/KrotofilC13} & \cite{DBLP:conf/acsac/KrotofilCML14} & \cite{lin2018tabor} & \cite{DBLP:conf/ccs/SarkarBM20} & \textbf{Our}\\
       \midrule
        PLC configuration & \checkmark & & & & & & & \checkmark & & \\
        HMI/WS configuration & & \checkmark & \checkmark & & & \checkmark & \checkmark & & & & \\
        Historian configuration & \checkmark & \checkmark & \checkmark & \checkmark & \checkmark & \checkmark & & & \checkmark & \checkmark &\\
        Network traffic & & & & \checkmark & \checkmark & & & & & & \\
        P\&ID & \checkmark & & \checkmark & & & \checkmark & \checkmark & \checkmark & \checkmark & \checkmark & \checkmark\\
       \bottomrule
  \end{tabular}
\end{table}

\section{Background}
\label{sec:background}

\subsection{Closed Control Loops}
\begin{figure}
  \centering
  \includegraphics[width=2.0in]{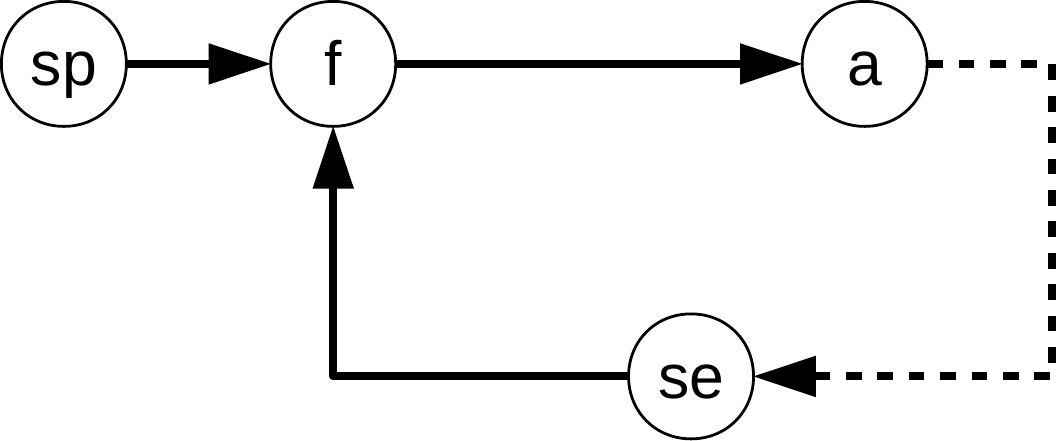}
  \caption{A CCL comprised of a setpoint (sp), a sensor (se), a control function (f), and an actuator (a). Solid lines represent communication in the \emph{cyber} domain whereas the dashed line
    represents interaction in the \emph{physical} realm.}
  \label{fig:simple_ccl}
\end{figure}

Closed Control Loops (CCLs) constitute a basic programming pattern for ICSs.
A CCL is comprised of four basic components, namely, a setpoint, a sensor, a control function, and an actuator (e.g., valve, heater, light, etc.).
The CCL's control function receives inputs from the environment through sensors, compares them with pre-established setpoints, and reacts with a compensatory action intended to minimize the difference.
The compensatory action is executed by an actuator in order to control the physical variable measured by the sensor (e.g., pressure, temperature, illumination, etc.).
Figure~\ref{fig:simple_ccl} depicts the simplest form of a CCL.

\vspace{-0.5cm}

The control function is a software component typically running in high availability embedded systems such as Programmable Logic Controllers (PLCs).
Control functions implement the control logic; for example, arithmetic operations, rate limiters, and other kinds of data processing functions.
From the function's perspective, hardware devices such as sensors and actuators are abstracted simply as variables to be read and written.
It is worth noting that in distributed control systems these variables might not necessarily reside on the same PLC.
Therefore, the communication to and from the control function might require network
transmissions.

\begin{figure}
    \centering
    \subfloat[\label{fig:multivariate}Cascade control.]
            {\includegraphics[width=1.5in]{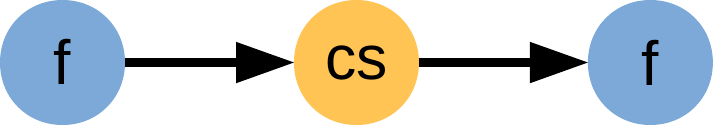}}\qquad
    \subfloat[\label{fig:common_actuator}Override control.]
            {\includegraphics[width=1.5in]{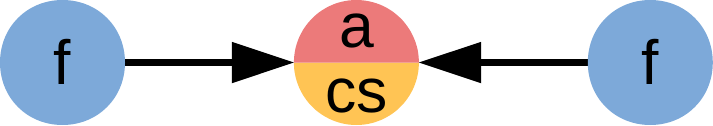}}\\
    \subfloat[\label{fig:common_setpoint}Shared setpoint.]
            {\includegraphics[width=1.5in]{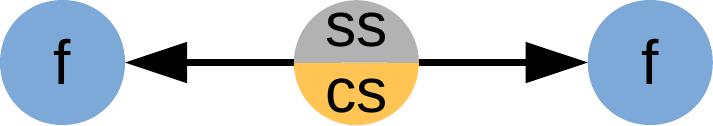}}\qquad
    \subfloat[\label{fig:common_sensor}Shared sensor.]
            {\includegraphics[width=1.5in]{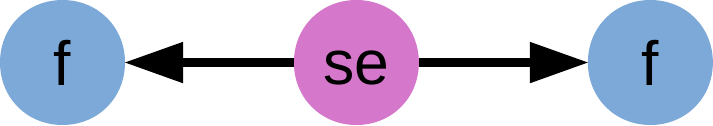}}\\
    \caption{Advanced CCL configurations. Color code: gray: static setpoint (ss), blue: controlling function (f), red: actuator (a), purple: sensor (se), and yellow: calculated setpoint (cs).}
    \label{fig:ccl_combinations}
\end{figure}

Advanced CCL configurations are often needed, e.g., to cope with system disturbances.
One of such configurations is called \emph{cascade control},
where one control function adjusts the setpoint of another control function~\cite{basicadvancedregulatory}.
This dynamically computed setpoint is called a \emph{calculated setpoint}.
In contraposition, we denote user-defined setpoints as \emph{static setpoints}.
Although we typically make an explicit distinction between static setpoints and calculated setpoints, in what follows, we use the word \emph{setpoint} to refer to either of them when such a distinction is irrelevant.
Graphically, an example of cascade control is shown in Figure~\ref{fig:multivariate}.

Another advanced CCL configuration is called \emph{override control}.
In this setting, one control function manipulates one variable during normal operation, however,
a second control function can take over during abnormal operation to prevent some safety, process, or equipment limit from being exceeded~\cite{basicadvancedregulatory}.
The notion of normal and abnormal operation is dependent on the physical process under control.
The variable under control by two or more control functions is typically used to manipulate one actuator or calculated setpoint.
Figure~\ref{fig:common_actuator} shows a graphical representation of the override control technique.

It is also common to find setpoints and sensors shared between two or more control functions, as shown in Figure~\ref{fig:common_setpoint} and Figure~\ref{fig:common_sensor}, respectively.
Shared setpoints provide a convenient centralized configuration for multiple control functions at once.
The motivation behind shared sensors is similar to that of shared setpoints,
which in addition reflects the fact that typically there are limited instances of physical sensors in the system under control.
An arbitrary number of the CCL configurations shown in Figure~\ref{fig:ccl_combinations} can be used to control ICSs~\cite{sharma2016overview}.

\vspace{-0.5cm}
\vspace{-1cm}

\subsection{Process Knowledge Data Sources}
Process comprehension is the main challenge attackers need to overcome to design successful and efficient attack strategies. Based on their capabilities, attackers can access different data sources of the targeted system, from collecting network captures to full access to an operator's workstation. 
Each data source provides a different view of the system. Depending on the attacker's interests, they could find some data sources more useful than others.
We follow the data source taxonomy presented in~\cite{DBLP:conf/ccs/GreenKA17} and describe the information that can be derived from them.
The data sources explored in their work include:

\paragraph{PLC Configuration.} PLCs are one of the most valuable components of ICSs. As they run the control logic that governs the process, they contain all the mechanisms to manipulate the system's physical properties. An attacker can learn the system's control logic and understand what sensors and network messages could jeopardize the system more quickly.

\paragraph{HMI/Workstation configuration.} HMIs are a valuable target for attackers because they typically run on Windows-based machines and have a comprehensive view of the system, as it was shown in the Ukraine power grid incident~\cite{UkranianBlackout2015}. Attackers with access to HMIs have a larger view of the system than individual PLCs. Attackers with access to HMIs could infer a more general view of the system, such as the type of industry they control~\cite{DBLP:conf/ccs/SarkarBM20}.

\paragraph{Historian configuration.} Historian data is also a valuable data source for attackers. As historians store previous traces of the system, they provide helpful information for attackers to understand the nominal behavior of the system. Historians are particularly valuable for model learning and the design of stealthy attacks~\cite{lin2018tabor}.

\paragraph{Network traffic.} Network flows describe how devices interact with each other. Attackers can sneak into the system's network and passively learn patterns of the system's behavior. Networks provide a source of realistic attack vectors that can drive the system into unsafe states~\cite{chen2019learning,chen2020active}.

\paragraph{Piping and Instrumentation Diagram (P\&ID).} P\&IDs show the functional relationship between the main components of the system, including piping, instrumentation and control devices. Attackers could learn valuable information about operational equipment and deduce which computing devices are worth being attacked.

\section{Identifying Near-Optimal Single-Shot Attacks}
\label{sec:identifying}

\subsection{System Model}
In this work, we focus on generic Industrial Control Systems (ICSs) implemented using a common programming pattern known as \emph{Closed Control Loop} (CCL).
The structure of CCLs is typically comprised of 4 components:
a control function, a setpoint, a sensor, and an actuator.
Control functions run in high availability embedded systems such as PLCs.
Setpoints, sensors, and actuators are abstracted as variables that serve as inputs and outputs for control functions.
The system might be distributed, meaning that the components of a single CCL reside on different PLCs but are interconnected through a communication network and exchange messages using standard ICS protocols (e.g., Modbus, EtherNet/IP, BACnet).

\subsection{Attacker Model}
\label{subsec:attacker_model}
\paragraph{Attacker Goal.} We consider an attacker that aims to lead to process damage (e.g., an emergency process shutdown) as fast as possible (as with longer attacks, the likelihood of detection rises). 

\paragraph{Attacker Capabilities.} To achieve this goal, the attacker will manipulate the values reported by a specific sensor (which simplifies the analysis, consistent with related work~\cite{DBLP:journals/ijcip/HuangCALTS09,DBLP:conf/nordsec/KrotofilC13}, and also limits the discussion to attacks on single sensors). Exactly how the sensor is manipulated is out of scope.
Prior work has demonstrated many ways to achieve this via physical-layer manipulations~\cite{DBLP:conf/ccs/TuRHRFH19}, traffic manipulations~\cite{DBLP:conf/nordsec/KrotofilC13}, or direct attacks on the PLCs or other hosts~\cite{abbasi2016ghost}.
In prior work, a number of approaches to select values to spoof have been discussed.
In this work we use the \emph{constant} and \emph{minimum/maximum} value attack from~\cite{DBLP:conf/ccs/CardenasALHHS11,DBLP:conf/sgcrc/UrbinaGTC16}.

\paragraph{Attacker Knowledge.} As obtaining information on the target process is costly (e.g., reconnaissance effort, bribes, industrial espionage), the attacker aims to minimize the type of information required to successfully run an attack. In particular, we assume an attacker that only has access to an abstract P\&ID. We will show later how from this diagram  a CCL graph can be derived.

\paragraph{Lack of Simulation.} In particular, the attacker does not have access to detailed physical process simulation environments. This implies (combined with the lack of detailed knowledge on the attacked process) that the effect of the attacker's manipulations cannot be reliably predicted in advance.

\subsection{Research Questions and Challenges}
Given this system and attacker model we address the following research question:\medskip

\noindent \emph{Given the set of all ICS sensors in the target system, how can the attacker identify the sensor to manipulate to achieve a near-optimal single-shot attack, with limited knowledge on the attacked process and system (i.e., a CCL graph)?}\medskip

Here we motivate the single-shot and near-optimal attack requirements in detail.

\paragraph{Need for Single-Shot Attacks.} While the system under attack does not use process-aware attack detection systems~\cite{DBLP:conf/sgcrc/UrbinaGTC16}, the human operators will eventually detect anomalous system conditions. Once anomalous system conditions are detected, a full-scale forensic investigation would be launched, which would remove the attacker's ability to launch further attacks. This means that the attacker will have to ideally perform an efficient attack on their first try. We call such attacks \emph{single-shot} attacks.

\paragraph{Near-Optimal Attacks.} Out of a large set of possible attacks within our attacker model, we expect there to be an attack that is optimal in the sense that its sensor manipulation leads to the fastest possible intended effect (i.e., shutdown of the process). But there might be other attacks that are nearly as fast. For the purpose of our evaluation, we introduce the concept of \emph{near-optimal} attacks. In particular, when ranking all possible attacks by their expected shutdown time, we call the three best attacks \emph{near-optimal}. We will also discuss alternative efficiency comparisons later in our discussion.

\paragraph{Challenge.} We note that prior ICS security research argues (or simply assumes) that \emph{detailed knowledge} about the system under attack is a requirement for a successful compromise~\cite{DBLP:conf/ccs/GreenKA17}. So the main challenge we will have to solve is to leverage the limited process knowledge (i.e., high-level CCL graph) to identify suitable sensors to attack --- without using attacker-side simulations to predict the outcome of process manipulations. If a near-optimal attack is reliably conducted on the first shot (even with limited knowledge on the system), our solution is considered successful.

\subsection{Identifying Near-Optimal Single-Shot Attacks in CCL Graphs}
We model ICSs as graph data structures that abstract the configuration of their closed control loops (CCLs).
The graph abstraction of a single ICS might consist of multiple subgraphs.
There are 5 types of nodes in our graphs that match the 5 components of CCLs:
\emph{static setpoints}, \emph{sensors}, \emph{control functions}, \emph{actuators}, and \emph{calculated setpoints}.

Formally, an ICS is modeled as a directed graph $G(V,E)$ where $V$ is a nonempty set of vertices (or nodes) and $E$ is a set of edges.
Every edge has exactly two vertices in $V$ as endpoints.
The direction of every edge $e \in E$ models the way information flows in the graph.
There are 5 partitions $SS$, $SE$, $F$, $A$ and $CS$ in $V$, that segregate the 5 types of nodes (\emph{static setpoint}, \emph{sensor}, \emph{control function}, \emph{actuator}, and \emph{calculated setpoint}, respectively), such that $SS \cup SE \cup F \cup A \cup CS = V$.
It is worth noting that although we assume knowledge about the type of the nodes (e.g., sensor), we do not require further details about them (e.g., temperature sensor, pressure sensor, etc.).

\begin{figure}
  \centering
  \subfloat[\label{fig:stepm-original}P\&ID of a chemical plant.]
           {\includegraphics[width=2.1in]{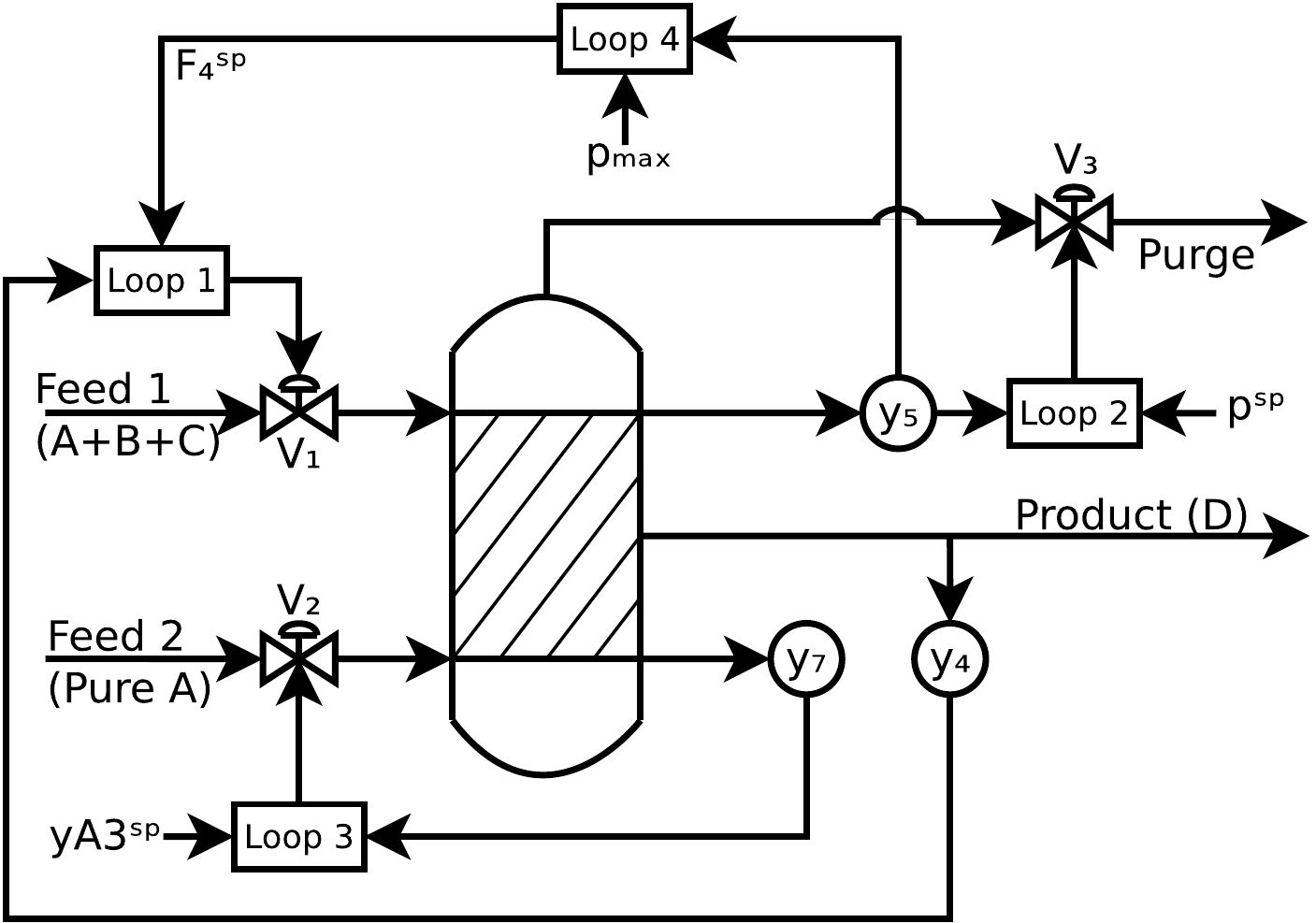}}
  \hspace{2em}
  \subfloat[\label{fig:stepm-graph}Graph abstraction of the control system derived from the P\&ID.]
           {\includegraphics[width=1.9in]{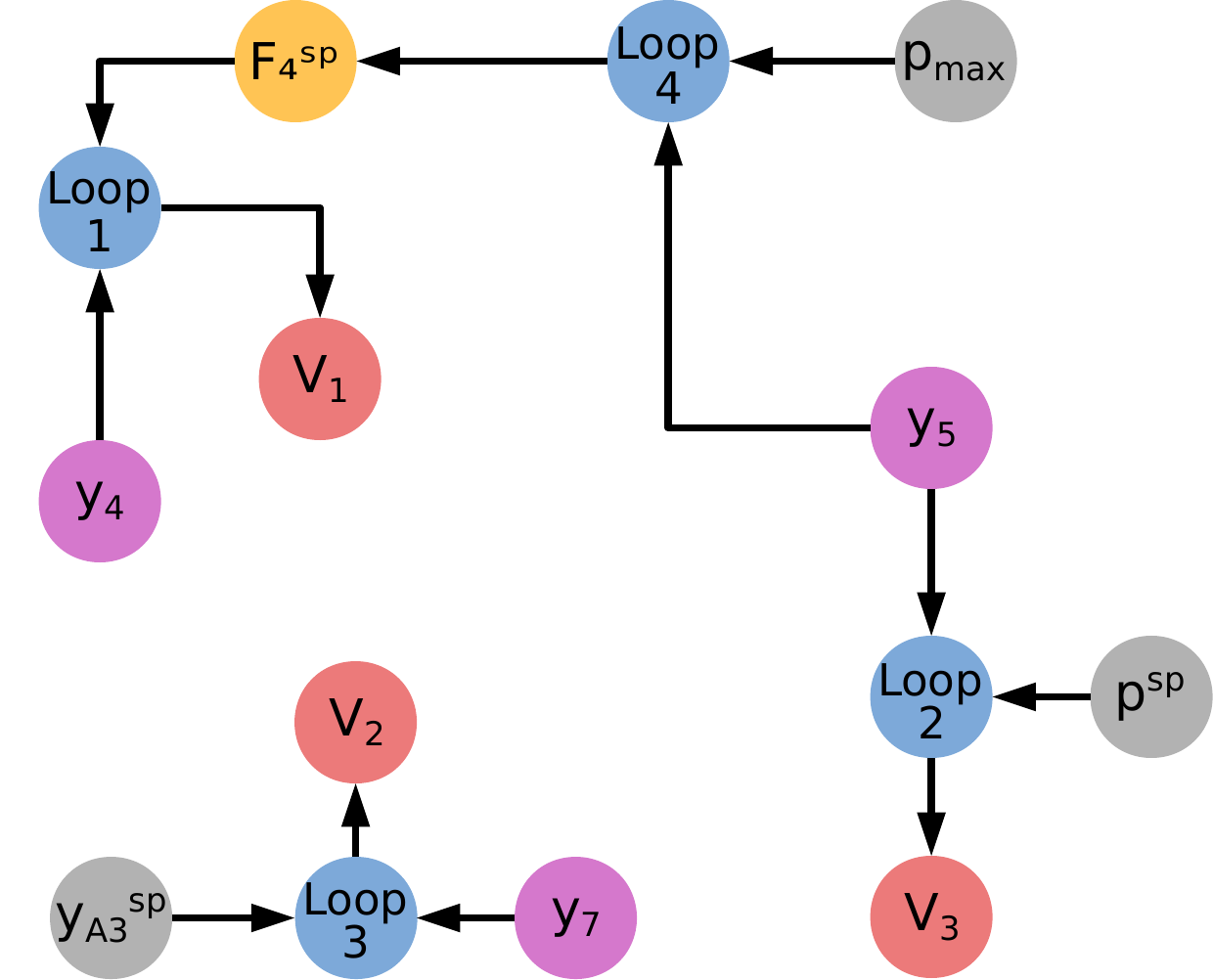}}
           \caption{Piping and instrumentation diagram (P\&ID) of an ICS and its corresponding graph abstraction.}
  \label{fig:stepm}
\end{figure}

There are different options to create CCL graphs.
It is possible to extract the CCL graph of ICSs from P\&IDs.
These diagrams show interconnected physical instruments and might contain information about the CCLs used in the ICS.
Figure~\ref{fig:stepm} shows an example of a P\&ID in Figure~\ref{fig:stepm-original},
and its corresponding CCL graph abstraction in Figure~\ref{fig:stepm-graph}.
Moreover, there are other options to create CCL graphs.
For instance, it is possible to create a CCL graph exclusively from network traffic in the BACnet protocol~\cite{dsnBACgraph}.
A use case from a real BACnet system is discussed and exemplified in Sect.~\ref{sec:discussion}.

\vspace{-0.5cm}

The proposed approach starts by searching for specific patterns in CCL graphs.
These patterns relate to well understood weaknesses originally analyzed in IT systems.
After that, a post-processing step is needed to filter out non-sensor targets.
This is important since according to our attacker model only sensor nodes will be targeted.
Finally, the attacker has to choose among the pre-selected sensors and has only one opportunity to compromise its integrity.
In the next sections, we explain how to execute near-optimal single-shot attacks and exemplify the proposed approach using the system presented in Figure~\ref{fig:stepm}.

\subsubsection{Pattern Matching}
\label{subsubsec:pattern_matching}
We propose to use the accumulated knowledge about software weaknesses in the IT domain and transfer it to the ICSs domain. In particular, we leverage Mitre's Common Weakness Enumeration (CWE) database, which specifies common IT weaknesses.
While we later use several of those CWE patterns in our implementation, here we focus on two entries which proved particularly useful experimentally. Our goal is to provide an intuitive comprehension of the proposed approach without limiting its applicability to a specific subset of weaknesses.
More formally, we identify each pattern $P_i$ with an index $i = 1,\dots,n$.
Each pattern matching query on the graph returns a subset $S_i \subseteq V$ of matching nodes using pattern $P_i$.\medskip

\paragraph{CWE-1108: Excessive Reliance on Global Variables.}
\emph{``The code is structured in a way that relies too much on using or setting global variables throughout various points in the code, instead of preserving the associated information in a narrower, more local context.''}~\cite{mitreCWE}.

Global variables are generally considered a bad software engineering practice.
Their main disadvantage is that malicious or benign-but-buggy changes to them will propagate and possibly disrupt many parts of the code.
Global variables can be observed in CCL graphs mainly due to shared setpoints and/or sensors.
As explained in Sect.~\ref{sec:background}, these are typical ways to combine CCLs in ICSs (see e.g., Figure~\ref{fig:common_setpoint} and Figure~\ref{fig:common_sensor}).

A suitable algorithm to identify global variables in CCL graphs is the \emph{out-degree} centrality.
This algorithm assigns a score to each node by counting their number of outgoing edges.
More formally, for every node $v \in V \setminus F$, the out-degree of $v$ is denoted as $d^{+}(v)$.
We explicitly disregard \emph{function} nodes in set $F$ since this particular weakness is exclusively about variables.
We select as potential targets those nodes whose value $d^{+}(v) \geq \tau$, for a context dependent threshold $\tau$.\medskip

\paragraph{CWE-1109: Use of Same Variable for Multiple Purposes.}
\emph{``The code contains a callable, block, or other code element in which the same variable is used to control more than one unique task or store more than one instance of data.''}~\cite{mitreCWE}.

Overloading a variable with multiple responsibilities might unnecessarily increase the complexity of the code around it.
Such complexity becomes an indirect security issue since it can hide potential vulnerabilities.

In control engineering, the usage of \emph{override controllers} deliberately creates a pattern in which two or more control functions manipulate a single variable (see Sect.~\ref{sec:background}, Figure~\ref{fig:common_actuator}).
The manipulated variable is commonly of type \emph{actuator} or \emph{calculated setpoint}.
Due to the widespread implementation of override controllers in ICSs, it is possible to find this pattern in real CCL graphs.

The automated identification of override controllers in a CCL graph can be done by computing the in-degree centrality of every node $v \in V \setminus F$.
This algorithm assigns a score to each node by counting their number of incoming edges.
As in the previous case, we explicitly disregard \emph{function} nodes in set $F$ since this particular weakness is exclusively about variables.
We denote the number of incoming edges of a node as $d^{-}(v)$.
Thus, we look at nodes whose $d^{-}(v) \geq \tau$, where typically $\tau = 2$.

\subsubsection{Post-processing}
According to our attacker model, it is a requirement to target nodes of type \emph{sensor} only.
Although some of the nodes that satisfy a weakness-related pattern might be of type \emph{sensor} other node types could be chosen too.
In fact, it is common to find \emph{setpoint} nodes as global variables and \emph{actuator} nodes as multi-purpose variables.

For each node $v \in S_i$, we first check whether $v \in SE$.
If $v$ is of type \emph{sensor} then $v$ is added to the result set $R_i \subseteq SE$.
Else, we search for every sensor $u \in SE$, such that there is a path from $u$ to $v$, and add $u$ to the result set $R_i$.
The pseudocode of this algorithm is listed in Alg.~\ref{alg:post-processing}.

\begin{algorithm}
  \caption{Pattern matching post-processing.}
  \label{alg:post-processing}
  \begin{algorithmic}
  \STATE input: set $S_i \subseteq V$ of nodes identified using $P_i$.
  \STATE output: set $R_i \subseteq SE$ of candidate sensor targets according to $P_i$.
  \STATE $R_i = \emptyset$
  \FORALL{$v \in S_i$}
    \IF{$v \in SE$}
    \STATE $R_i = R_i \cup \{v\}$
    \ELSE
    \FORALL{$u \in SE$}
        \IF{exist\_path(from:$u$, to:$v$)}
        \STATE $R_i = R_i \cup \{u\}$
        \ENDIF
    \ENDFOR
    \ENDIF
  \ENDFOR
  \end{algorithmic}
\end{algorithm}
\vspace{-1cm}

\subsubsection{Single-Shot Attacks}
\label{subsubsec:1shot}
After the pattern matching phase, the attacker has to choose one target from all the result sets obtained.
To choose one target from $\cup_{i=1}^{n}R_{i}$, we assign each sensor a score depending on how many result sets $R_i$ they occur in.
The reasoning behind this scoring system is that the sensors that have been identified by more weakness-related patterns have a greater chance of attack success.
This selection criteria creates a subset of targets $T \subseteq \cup_{i=1}^{n}R_{i}$ among which only one has to be selected.
Without further insights about the targeted infrastructure besides the CCL graph, it is hard to provide a meaningful sensor selection strategy from $T$.
However, we hypothesize that $T$ (1)~will be smaller that the original sensor set; and (2)~will contain only sensors capable of causing near-optimal shutdown times (SDTs).
A smaller subset to choose sensors from, gives the attacker a probabilistic advantage over, e.g., an attacker that has to choose a sensor from the whole sensor set $SE$.
Moreover, since we hypothesize that \emph{all} sensors in $T$ are capable of causing near-optimal SDTs, a simple (e.g., random) selection strategy is suitable from the attacker's perspective.

To compromise the integrity of a sensor, the attacker overwrites legitimate sensor readings in such a way that all linked control functions will take the attacker's desired value instead.
This value is fixed throughout the whole period of the attack,
however, the attacker is free to choose the value at will.

\subsection{Motivating Example}
\label{subsec:motivating_example}
Figure~\ref{fig:stepm-original} shows the model of a chemical plant originally described in~\cite{ricker1993model}.
Four chemical components referred to as A, B, C, and D, are part of the process.
The first three components are combined in the reactor to create the final product D.
The goal of the software controlling the plant is to keep a stable high quality production rate while minimizing the waste of raw material.
There are four control functions that take as input the values coming from three sensors and four setpoints (1 calculated setpoint and 3 static setpoints).
The output of the functions aims to control three valves in the plant.

From a security perspective, there are several details of the plant that a potential attacker would like to know to execute an attack.
For example, the types of sensors (i.e., pressure, temperature), the chemical reaction carried out by the plant (i.e., $A + C \xrightarrow[]{B} D$), the maximum pressure supported by the reactor (3,000 kPa), etc.
Detailed process knowledge has been used in previous works to exemplify simulated attacks against this chemical plant~\cite{DBLP:conf/ccs/CardenasALHHS11,DBLP:journals/ijcip/HuangCALTS09}.
Our approach, however, assumes limited process knowledge.
More precisely, we only assume access to the CCL graph of the targeted infrastructure (Fig.~\ref{fig:stepm-graph}), which lacks all of the previously mentioned details.

Figure~\ref{fig:stepm-graph} depicts the CCL graph of the chemical plant, which can be easily derived from its piping and instrumentation diagram.
This graph shows two kinds of CCL combinations: (1) a shared sensor $y_5$ between control functions \emph{Loop 4} and \emph{Loop 2};
and (2) a cascade control in which control function \emph{Loop 4} sets a calculated setpoint $F_{4^{sp}}$ to control function \emph{Loop 1}.

The proposed approach looks for weakness patterns in the chemical plant's CCL graph that could identify suitable sensors to target.
The search for multi-purpose variables (functions excluded) aims at nodes whose in-degree is greater or equal than two; no results are produced in this case.
The search for global variables (functions excluded) looks for nodes whose out-degree is greater than a predefined threshold $\tau$.
All setpoints have an out-degree of one which makes them unfit to be labeled as global variables.
However, the three sensors $y_4$, $y_5$, and $y_7$ have an out-degree of 1, 2, and 1, respectively.
This small sensors sample shows an average out-degree of 1.33 with standard deviation of 0.58.
Defining $\tau$ as the sum of the average and one standard deviation ($\tau = 1.91$) sets node $y_5$ as a potential target according to the proposed approach.
The result set would have sensor $y_5$ as an ideal candidate to target in this particular infrastructure.
This result is concordant with previous works which state that ``[i]n general we found that the plant is very resilient to attacks on $y_7$ and $y_4$... If the plant operator only has enough budget to deploy advanced security mechanisms for one sensor (e.g., tamper resistance, or TPM chips), $y_5$ should be the priority''~\cite{DBLP:conf/ccs/CardenasALHHS11}.

\section{Implementation}
\label{sec:implementation}
We implemented the proposed approach  on top of Neo4j version 4.1.2~\cite{neo4j}.
Neo4j is a noSQL database engine specialized in graph data structures.
Neo4j offers a natural way to store CCL graphs and a high level query language that allows us to perform complex queries in just a few lines of code.

\begin{figure}
  \centering
  \includegraphics[width=3.0in]{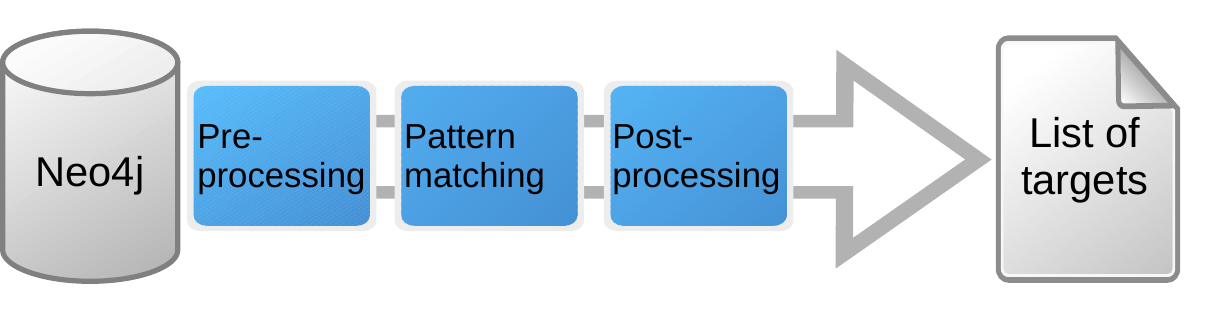}
  \caption{Implementation of the proposed approach.}
  \label{fig:implementation}
\end{figure}

We assume that the attacker already has access to the CCL graph of the targeted infrastructure and has persisted it in a Neo4j database.
Without any further knowledge, the selection of the sensors to attack is done through a set of queries on the graph.
Such queries must be written in Neo4j's query language called \emph{Cypher Query Language}.
Figure~\ref{fig:implementation} depicts the overall process to obtain a subset $R_i$ of candidate sensor targets according to a pattern $P_i$.

Our implementation includes a \emph{pre-processing} stage that pre-computes information required in the \emph{pattern matching} phase.
Since both weakness-related patterns discussed in Sect.~\ref{subsubsec:pattern_matching} are based on the in- and out-degree centrality algorithms, the \emph{pre-processing} stage consists of queries that assign the in- and out-degree to every node in the graph.
Both queries are shown in Appendix~\ref{appendix:cql} code listing~\ref{lst:degree}.

The \emph{pattern matching} phase consists in finding nodes that satisfy a specific condition in the graph's topology.
These particular conditions pinpoint nodes of interest from the attacker's perspective.
For the sake of brevity, here we discuss two weaknesses, namely, global variables and multi-purpose variables.
Global variables are nodes whose out-degree is greater than a context dependent threshold $\tau$.
In our implementation, we define $\tau$ as the average plus one standard deviation of the out-degree of nodes segregated per type.
Code listing~\ref{lst:global}, in Appendix~\ref{appendix:cql}, shows the query for the global variable pattern.

The second \emph{pattern matching} query looks for multi-purpose variables (e.g., override controllers).
This pattern is simpler since we only need to find nodes with an in-degree greater or equal than 2.
Code listing~\ref{lst:multi}, in Appendix~\ref{appendix:cql}, shows the query used to find the multi-purpose variable pattern.

Lastly, the \emph{post-processing} phase replaces non-sensor nodes found during the \emph{pattern matching} phase, with sensor nodes that have a path to them,
thus, influencing their behavior.
This ensures that the list of targets for each weakness-related pattern is comprised exclusively of sensor nodes.
Code listing~\ref{lst:path} (Appendix~\ref{appendix:cql}) shows the main task of the \emph{post-processing} query.
This query looks for all sensor nodes (src) that have a path to the node of interest (dest), regardless of the length of the path.

\vspace{-0.5cm}

\vspace{-1cm}

\section{Experimental Evaluation}
We evaluate the proposed approach in a realistic industrial control system.
We use this environment to perform experimental attacks against \emph{all} relevant sensors to obtain a ground-truth about the severity of the attacks in terms plant availability.
Due to the large number of attacks required to obtain the ground-truth, we opted for a simulated environment where the attacks can be executed without safety concerns.
The simulated plant is known as the Tennessee Eastman Plant and has been extensively used in previous cybersecurity research~\cite{DBLP:conf/ccs/CardenasALHHS11,DBLP:journals/ijcip/HuangCALTS09,DBLP:conf/nordsec/KrotofilC13,DBLP:conf/acsac/KrotofilCML14,DBLP:conf/ccs/SarkarBM20}.

\subsection{Tennessee Eastman Plant}
The seminal 1993 paper by Downs and Vogel describes the Tennessee Eastman Plant (TEP)~\cite{downs1993plant}.
Their description includes, among other details, the expected input and output of the plant, each step of the process from start to end, and the hardware available to control the process.
The control hardware includes 41 sensors and 12 actuators, depicted in Appendix~\ref{appendix:tep_diagram} (Figure~\ref{fig:tep}).
Their paper describes 20 disturbances commonly found in real chemical plants.
For instance, sticky valves, changes in chemical reaction kinetics, and random variations in the composition of input streams.
Each disturbance has a unique numeric identifier in the range [1--20].
Finally, there are process operating constraints (e.g., maximum reactor temperature, maximum reactor pressure, etc.) that must be satisfied at all times or the plant shuts down.
Downs and Vogel present the challenge of implementing a control strategy for the TEP.
To ease engagement in the challenge, the authors provide software source code to simulate the core components of the TEP such as the sensors, actuators, disturbances, and process operating constraints, leaving space for the missing control strategy.

Several authors have proposed control strategies for the TEP~\cite{larsson2001self,luyben1999plantwide,lyman1992plant,ricker1996decentralized}.
The main difference between control strategies are the robustness against external disturbances, the optimization objectives, and the mechanisms to set the production rate.
The TEP challenge is considered an open-ended problem without a unique correct solution~\cite{luyben1999plantwide}.

In this work, we perform an extensive analysis of two control strategies for the TEP in terms of plant \emph{availability} upon sensor integrity attacks.
The first strategy, proposed by Larsson et al.~\cite{larsson2001self},
is available in the MATLAB/Simulink environment~\cite{rickerMatlabArchive}\footnote{Simulation and results available at \url{https://gitlab.com/eastman_tennessee/larsson}.}.
The second strategy, proposed by Luyben et al.,
is available in the Fortran programming language~\cite{luyben1999plantwide}.
We translated it to the MATLAB/Simulink environment and published it\footnote{Simulation and results available at \url{https://gitlab.com/eastman_tennessee/luyben}.}.\medskip

\subsection{Experimental Attacks}
All simulations are executed on the MATLAB/Simulink environment.
Specifically, we use MATLAB version R2015a running on Windows 10.
The simulations are configured to run for 72 hours under attack.
However, some of the attacks cause violations in the process operating constraints.
As a consequence, some simulations stop earlier than expected.
We refer to the time elapsed since the beginning of the attack and until the simulation stops as the shutdown time (SDT).

An \emph{experiment} is a set of simulations using the same environmental conditions (i.e., disturbances) and attack strategy.
In our experiments, we execute simulations with and without disturbances.
The disturbances considered are those in the range [1--13] and are executed one at a time.
According to the original TEP paper, disturbances in the range [14--20] should be used in conjunction with other disturbances~\cite{downs1993plant}.
The combinatorial explosion of such constraint deters us from executing simulations with disturbances in the range [14--20].
The attack strategy is the way in which the attacker chooses the value used to compromise the integrity of sensors.
We use three different attack strategies.
First, assuming that the attacker does not have any knowledge about the targeted sensor, we choose the constant 127.
This number is small enough to fit in 1 byte (signed int) which ensures that most industrial protocols will deliver the malicious value in a single packet, thus, executing a stealthy attack.
For the second and third attack strategies, we assume that the attacker knows historic sensor readings, in which case we choose the minimum and maximum values observed per sensor.
Although this additional knowledge is not required by our approach, we use these attack strategies to
compare our results with previous works.

After the experimental setup has been defined, we execute the attacks against each sensor, one at a time, and record its SDT (or 72 hours if no shutdown happens).
If a simulation finishes at 72 hours, we assume that the attack does not cause a shutdown.
The experiment finishes when all sensors used by the control strategy have been attacked.
We rank the targets according to their SDT to identify the fastest SDT in the experiment and the sensor that causes it.
We refer to them as the \emph{optimal SDT} and the \emph{optimal target}, respectively.
In general, we refer to the top-3 fastest attacks as \emph{near-optimal attacks}.

To put our results in perspective, we compare the chances of achieving a near-optimal attack of an attacker using our approach (with the capabilities described in Sect.~\ref{subsec:attacker_model}) and an attacker who picks, uniformly at random, one sensor to target
(we assume that both attackers use the same attack strategy).
Hereafter, we refer to the latter simply as the \emph{random attacker}.

\subsubsection{Control Strategy \#1}
The control strategy by Larsson et al.~\cite{larsson2001self},
uses only 9 actuators and 16 sensors out of the 12 actuators and 41 sensors available in the TEP.
Additionally, there are 20 control functions, 9 static setpoints, and 12 calculated setpoints.
The CCL graph of this control strategy is comprised of 66 nodes divided in 3 subgraphs.
An illustration of the graph is shown in Appendix~\ref{appendix:control_strategy_1} (Figure~\ref{fig:Larsson_graph}).

We use the queries detailed in Sect.~\ref{sec:implementation} to identify potential targets.
First, we compute the in- and out-degree centrality for all the sensors, setpoints, and actuators.
Then, we run the pattern matching queries that aim at finding global variables and multi-purpose variables.
The first query identifies the calculated setpoint number 12 (located in the middle of the largest subgraph) as a \emph{global variable}.
Although there are 16 sensors used in this control strategy, the post-processing phase identifies that only one of them has a path to the global variable: \emph{sensor 17} (i.e., $R_1=\{17\}$).
Such a path can be visually confirmed following the direction of the edges in Figure~\ref{fig:Larsson_graph}.
No \emph{multi-purpose variables} are identified in this control strategy (i.e., $R_2=\emptyset$)
Thus, the final target set $T=\{17\}$ contains only one sensor, which makes the target selection easier for a single-shot attack.

We hypothesize that sensor 17 is a near-optimal target for control strategy \#1.
To test our hypothesis we execute 28 experiments comprised of 448 individual simulations that account for 11,047.465 simulated hours ($\sim$1.26 years).
For the first two experiments we use the constant value attack strategy.
In one of the experiments we set ideal environmental conditions (no disturbance) and in the other experiment we enable disturbance \#8.
We choose disturbance \#8 because previous works have used exclusively this disturbance for their experiments.
The results of the first two experiments are shown in Table~\ref{tab:Larsson_127}.
Regardless of the environmental conditions, the results are consistent in the top half of the table with greater variations in the bottom half.
In these two experimental settings, a \emph{random attacker} would have 1/16 chances ($\approx$ 6\%) of choosing sensor 40 (which does not cause a shutdown), and 3/16 ($\approx$ 18\%) of choosing a near-optimal target.
Considering only the 15 sensors that do cause a shutdown, a \emph{random attacker} would achieve
an average SDT of 2.66 hours without disturbance and 2.52 hours under disturbance \#8.
On the other hand, an attacker using our approach has to choose one sensor from $T=\{17\}$, thus, ensuring 100\% chance of finding a near-optimal target with a SDT of 0.192 hours in both cases ($\approx$ 11 minutes) and a difference of about 5 minutes behind the optimal target (sensor 9 with SDT of 0.101 hours.).

Previous works have used the minimum and maximum value attack strategies against this control strategy.
Unlike previous works that have used only one environmental condition to execute their experiments (disturbance \#8), we execute our experiments under 13 different environmental conditions (including disturbance \#8) to gather data from more diverse scenarios and gain more confidence in our results.
Disturbance \#6 is excluded because it is not supported by this particular control strategy, which means that a shutdown happens even without any attacks~\cite{larsson2001self}.
Due to space constraints, we summarize our results per attack strategy, which shows the average SDT and standard deviation for each sensor among all the simulations.
The results, detailed in Tables~\ref{tab:Larsson_sensor_positions_min} and~\ref{tab:Larsson_sensor_positions_max}, show that sensor 17 is a near-optimal target with an average SDT of 1.21 hours and 1.07 hours, respectively.
For the minimum value attack strategy, a random attacker would achieve an average SDT of 24.23 hours,
whereas in the maximum value attack strategy an average SDT of 24.87 (excluding sensor 1).
Finally, for both the minimum and maximum attack strategies, sensor 17 is ranked in the second position.
This confirms that sensor 17 is a near-optimal target not only during specific plant conditions, but in many different situations.

\begin{table}
  \begin{center}
    \caption{Experiments using the constant value attack strategy under two different environmental conditions. Sensor 17, identified as a near-optimal target, is highlighted.}
    \label{tab:Larsson_127}
    \begin{tabular}{c|c||c|c}
      \multicolumn{2}{c}{No disturbance} & \multicolumn{2}{c}{Disturbance 8}\\
      \toprule
      Sensor & SDT (h) $^{\blacktriangle}$ & Sensor & SDT (h) $^{\blacktriangle}$\\
      \midrule
      9 & 0.101 & 9 & 0.101\\
      14 & 0.181 & 14 & 1.181\\
      \rowcolor[gray]{.8}17 & 0.192 & 17 & 0.192\\
      \hline
      11 & 0.426 & 11 & 0.427\\
      8 & 0.431 & 8 & 0.430\\
      4 & 0.526 & 4 & 0.526\\
      31 & 0.557 & 31 & 0.560\\
      12 & 0.569 & 12 & 0.567\\
      3 & 1.604 & 3 & 1.607\\
      2 & 1.757 & 2 & 1.614\\
      15 & 2.126 & 15 & 2.052\\
      1 & 7.335 & 5 & 5.644\\
      5 & 7.609 & 7 & 6.971\\
      10 & 8.182 & 10 & 7.249\\
      7 & 8.297 & 1 & 9.691\\
      40 & 72 (no shutdown) & 40 & 72 (no shutdown)\\
      \bottomrule
    \end{tabular}
  \end{center}
\end{table}

\begin{table}
    \begin{minipage}{.46\textwidth}
    \caption{Summary of experiments considering the minimum value attack strategy under 13 different environmental conditions. Sensor 17, identified as a near-optimal target, is highlighted.}
    \label{tab:Larsson_sensor_positions_min}
    \begin{tabular}{c|c|c}
      \toprule
      Sensor & Avg. SDT (h) $^{\blacktriangle}$ & Std. Deviation\\
      \midrule
                               4  & 0.79  & 0.67\\
     \rowcolor[gray]{.8}       17 & 1.21  & 0.76\\
                               9  & 2.06  & 0.73\\
                               \hline
                               8  & 2.66  & 0.41\\
                               3  & 4.11  & 0.35\\
                               2  & 4.55  & 0.75\\
                               7  & 7.70  & 2.19\\
                               12 & 7.97  & 2.25\\
                               14 & 8.94  & 2.39\\
                               15 & 11.18 & 3.96\\
                               5  & 14.53 & 17.49\\
                               31 & 55.15 & 27.74\\
                               11 & 66.64 & 19.31\\
                               40 & 66.69 & 19.14\\
                               10 & 66.73 & 19.02\\
                               1  & 66.78 & 18.82\\
     \bottomrule
    \end{tabular}
    \end{minipage} 
\hfill
    \begin{minipage}{.46\textwidth}
    \caption{Summary of experiments considering the maximum value attack strategy under 13 different environmental conditions. Sensor 17, identified as a near-optimal target, is highlighted.}
    \label{tab:Larsson_sensor_positions_max}
    \begin{tabular}{c|c|c}
      \toprule
      Sensor & Avg. SDT (h) $^{\blacktriangle}$ & Std. Deviation\\
      \midrule
                               9  & 0.57  & 0.19\\
     \rowcolor[gray]{.8}       17 & 1.07  & 0.14\\
                               4  & 1.27  & 0.21\\
                               \hline
                               3  & 2.60  & 0.42\\
                               8  & 2.83  & 0.46\\
                               2  & 3.32  & 0.66\\
                               12 & 5.44  & 1.36\\
                               14 & 8.79 & 2.45\\
                               15 & 11.10 & 4.36\\
                               5  & 12.64 & 17.91\\
                               31 & 56.78 & 27.62\\
                               11 & 66.59 & 19.50\\
                               10 & 66.64 & 19.34\\
                               40 & 66.70 & 19.10\\
                               7  & 66.73 & 19.00\\
                               1  & 72 (no shutdown) & 0.00\\
     \bottomrule
    \end{tabular}
    \end{minipage}
\end{table}

\subsubsection{Control Strategy \#2}
The second control strategy, proposed by Luyben et al.~\cite{luyben1999plantwide}, uses 10 sensors and 10 actuators from those available in the plant.
This control strategy requires 13 static setpoints, 13 control functions, and 1 calculated setpoint.
In total, the CCL graph is comprised of 47 nodes.
The overall CCL graph of the system (depicted in Figure~\ref{fig:Luyben_graph} in Appendix~\ref{appendix:control_strategy_2}), has 6 subgraphs.

As in the previous control strategy, we use the queries detailed in Sect.~\ref{sec:implementation} to identify sensor targets, starting with the computation of the in- and out-degree centrality for all the sensors, setpoints, and actuators.
The query regarding global variables identifies sensors 8, 12, and 15 (i.e., $R_1=\{8,12,15\}$).
The query regarding multi-purpose variables identifies actuators 1, 2, 7, and 11.
The post-processing phase looks for sensor nodes that have a path to these actuators and identifies sensors 8, 12, 15, and 29 (i.e., $R_2=\{8,12,15,29\}$).
As described in Sect.~\ref{subsubsec:1shot}, the final targets subset $T$ contains those sensors identified by most weakness-related patterns.
In this particular case, $T=\{8,12,15\}$ because these sensors occur in both $R_1$ and $R_2$.
Finally, the attacker selects one target $t \in T$ at random.

We hypothesize that sensors 8, 12, and 15 are near-optimal targets against control strategy \#2.
To test our hypotheses we execute 30 experiments comprised of 300 individual simulations that account for 11,079.163 simulated hours ($\sim$1.26 years).
As in the previous control strategy, we begin with two experiments using the constant value attack strategy under two different environmental conditions.
The first experiment without any disturbance and the second experiment under disturbance \#8.
The results of the first two experiments, detailed in Table~\ref{tab:Luyben_127}, show no significant differences between both plant conditions.
In these environmental conditions, the \emph{random attacker} has a 30\% chance of choosing a near-optimal target (3 out of 10 sensors), and the same chances of choosing a sensor that do not cause a shutdown.
On the other hand, an attacker using our approach has a  66\% chance of finding a near-optimal target (2 out of 3 sensors) since sensors 12 and 15 are near-optimal but not sensor 8.
The \emph{random attacker} can achieve an average SDT of 1.59 hours without disturbance and 1.51 hours under disturbance \#8 (excluding the 3 sensors that do not cause a shutdown).
On the other hand, an attacker using our approach achieves an average SDT of about 0.50 hours in both cases; more than 1 hour faster.

As for control strategy \#1, we execute additional experiments using the minimum and maximum attack strategies.
This time we use all disturbances in the range [1--13] because this control strategy is able to handle all of them.
Thus, we execute two attack strategies over 14 environmental conditions (no disturbance + 13 disturbances), which adds up to 28 additional experiments.
Again, due to space constraints, we summarize our results per attack strategy. 
Tables~\ref{tab:Luyben_sensor_positions_min} and~\ref{tab:Luyben_sensor_positions_max} show the average SDT and standard deviation for each sensor attack throughout all the simulations.
As in the first two experiments, for the minimum and maximum attacks, the \emph{random attacker} has 30\% chance to find a near-optimal target but also 30\% chance of finding a target that does not cause a shutdown.
For our attacker, however, the chances of finding a near-optimal target differ per attack strategy.
The minimum value attack gives our attacker only 1 out of 3 chances of finding the near-optimal sensor 15 (33\%); a marginal benefit with respect to the \emph{random attacker}.
However, our attacker does not have the 30\% risk of a no-shutdown attack.
After excluding the 3 sensors that do not cause a shutdown, the average SDT of the \emph{random attacker} is 20.75 hours; significantly larger than the average SDT of 4.4 hours for our attacker.
For the maximum value attack, our attacker has 2 out of 3 chances of picking a near-optimal target
After excluding the 3 sensors that do not cause a shutdown, the average SDT of the \emph{random attacker} is 25.96 hours.
On the contrary, an attacker using our approach achieves an average SDT of 1.96 hours.

\begin{table}
  \begin{center}
    \caption{Experiments using the constant value attack strategy under two different environmental conditions. The 3 sensors identified as near-optimal targets are highlighted.}
    \label{tab:Luyben_127}
    \begin{tabular}{c|c||c|c}
      \multicolumn{2}{c}{No disturbance} & \multicolumn{2}{c}{Disturbance 8}\\
      \toprule
      Sensor & SDT (h)  $^{\blacktriangle}$ & Sensor & SDT (h)  $^{\blacktriangle}$\\
      \midrule
      7 & 0.144 & 7 & 0.144\\
      \rowcolor[gray]{.8}15 & 0.161 & 15 & 0.161\\
      \rowcolor[gray]{.8}12 & 0.311 & 12 & 0.311\\
      \hline
      9 & 0.440 & 9 & 0.440\\
      \rowcolor[gray]{.8}8 & 1.031 & 8 & 1.032\\
      11 & 2.486 & 11 & 2.647\\
      23 & 6.549 & 23 & 5.844\\
      18 & 72 (no shutdown) & 18 & 72 (no shutdown)\\
      29 & 72 (no shutdown) & 29 & 72 (no shutdown)\\
      30 & 72 (no shutdown) & 30 & 72 (no shutdown)\\
      \bottomrule
    \end{tabular}
  \end{center}
\end{table}
\vspace{-1.0cm}

\begin{table}
\begin{minipage}{.46\textwidth}
    \caption{Summary of experiments considering the minimum value attack strategy under 14 different environmental conditions. The 3 sensors identified as near-optimal targets are highlighted.}
    \label{tab:Luyben_sensor_positions_min}
    \begin{tabular}{c|c|c}
      \toprule
      Sensor Id. & Avg. SDT (h) $^{\blacktriangle}$ & Std. Deviation\\
      \midrule
      \rowcolor[gray]{.8}      15 & 0.65  & 0.20\\
                               9  & 1.28  & 0.51\\
                               7  & 1.72  & 0.79\\
      \hline
      \rowcolor[gray]{.8}      12 & 4.25  & 1.72\\
      \rowcolor[gray]{.8}      8  & 8.29  & 1.98\\
                               23 & 63.03 & 22.83\\
                               30 & 66.01 & 16.82\\
                               29 & 72 (no shutdown) & 0.00\\
                               11 & 72 (no shutdown) & 0.00\\
                               18 & 72 (no shutdown) & 0.00\\
      \bottomrule
    \end{tabular}
    \end{minipage}
\hfill
\begin{minipage}{.46\textwidth}
    \caption{Summary of experiments considering the maximum value attack strategy under 14 different environmental conditions. The 3 sensors identified as near-optimal targets are highlighted.}
    \label{tab:Luyben_sensor_positions_max}
    \begin{tabular}{c|c|c}
      \toprule
      Sensor Id. & Avg SDT (h) $^{\blacktriangle}$ & Std. Deviation\\
      \midrule
      \rowcolor[gray]{.8}      12 & 0.69  & 0.16\\
      \rowcolor[gray]{.8}      15 & 0.70  & 0.07\\
                               9  & 1.09  & 0.40\\
      \hline
      \rowcolor[gray]{.8}      8  & 4.48  & 0.55\\
                               23 & 52.01 & 29.46\\
                               7  & 55.30 & 27.60\\
                               29 & 67.46 & 16.97\\
                               30 & 72 (no shutdown)    & 0.00\\
                               11 & 72 (no shutdown)    & 0.00\\
                               18 & 72 (no shutdown)    & 0.00\\
      \bottomrule
    \end{tabular}
    \end{minipage} 
\end{table}

\section{Discussion}
\label{sec:discussion}
\paragraph{CCL Graphs.} The first challenge faced by graph-based studies is the creation of the graph itself.
There are different possibilities to create CCL graphs like those used in this work.
Piping and Instrumentation Diagrams (P\&IDs) are a suitable alternative because ICS documentation can be obtained through a variety of illegal means (e.g.,~phishing, social engineering, bribes) or simply downloaded from public repositories~\cite{DBLP:conf/latw/KonstantinouSM16}.
The creation of a CCL graph from a P\&ID could even be automated using diagram digitization techniques such as~\cite{patentArroyo}.

Another way to create CCL graphs leverages the rich semantics of some industrial communication protocols.
A concrete example is the BACnet protocol (ISO 16484-5), commonly used to automate diverse services in hospitals, airports, and other buildings~\cite{ashrae_bacnet_standard}.
In these environments, the CCL programming pattern is extensively used.
For that reason, the BACnet protocol implements an application layer object called \emph{Loop}, which eases the implementation of CCLs.
This object contains properties that point to other BACnet objects abstracting CCL components such as sensors, setpoints, and actuators.
Since BACnet objects are regularly exchanged through the network, it is possible to create CCL graphs in a fully automated way simply by sniffing the traffic~\cite{dsnBACgraph}.
We used this method to create the CCL graph of a real BACnet system comprised of more that 20 buildings located at the University of Twente.
We can confirm that this method is capable of creating large CCL graphs (4,771 nodes) just by passively listening the network traffic.
In Appendix~\ref{appendix:BACnet} we show three subgraphs obtained during our analysis and a plot of the node discovery rate during our 7-week long experiment (Figures~\ref{fig:real_bas} and~\ref{fig:discovery_rate}, respectively).
In general, we note similar structures in the BACnet graphs and the TEP graphs, which suggests the possibility to apply the proposed approach to other systems besides ICSs.

\paragraph{Additional CWE Weaknesses.} To describe the proposed approach, we elaborate on two weaknesses from Mitre's CWE database, in particular, global variables (CWE-1108) and multi-purpose variables (CWE-1109)~\cite{mitreCWE}.
We emphasize these two because they proved particularly useful to find weaknesses in both TEP implementations analyzed in our evaluation.
Although it is not our goal to provide an exhaustive list of software weaknesses that can be mapped to ICSs,
here we discuss additional weaknesses that could be observed in CCL graphs.

Circular dependencies (CWE-1047) happen when \emph{``[t]he software contains modules in which one module has references that cycle back to itself.''}~\cite{mitreCWE}.
In ICSs, circular dependencies can occur, for example, through control functions that write their output to a calculated setpoint node that, in turn, is the input of another control function node and so on, until at some point the sequence of references return to the initial node.
Although none of the CCL graphs analyzed showed a circular dependency pattern, we found a similar structure in the BACnet system previously discussed.
We illustrate it in Appendix~\ref{appendix:BACnet} Figure~\ref{fig:act_set}.

Deep nesting (CWE-1124) manifests in software that \emph{``contains a callable or other code grouping in which the nesting / branching is too deep.''}~\cite{mitreCWE}.
The software implementation of ICSs might contain a deep nesting weakness whenever a long sequence of closed control loops are chained together. For example, several cascade controllers concatenated.
In this setting, the precise definition of `long sequence' is determined by a context dependent threshold.

\paragraph{Countermeasures.} The most promising way to thwart our described approach to identify near-optimal single-shot attacks would be by preventing the attacker’s ability to gain knowledge of the CCL graph representation of the targeted infrastructure in the first place. However for many ICSs, P\&IDs are publicly available, which an attacker can use to generate a corresponding CCL graph as we described before. Therefore, it would be necessary to keep such P\&IDs secret and to protect them from being leaked to attackers. We stress here that further protective measures might be needed to prevent an attacker from being able to generate the CCL graph through possibly other information such as, in the case of building automation systems, using the BACnet protocol, for which eavesdropping on the  traffic is already sufficient to automatically generate CCL graphs~\cite{dsnBACgraph}.

Our approach relies on identifying specific weakness patterns in the CCL graph which can be matched with well-known software weaknesses in the IT domain (e.g., via Mitre’s CWE database). Therefore, if a CCL graph does not contain any of these matching weaknesses, then our approach would fail in identifying near-optimal single-shot attacks with a better probability than by simply picking any sensor from the set of all sensors at random. This means that if an attacker cannot be prevented from creating the CCL graph, then another way to thwart our method would be by ensuring that the CCL graph contains no weakness patterns. One option to achieve this would be by performing a re-engineering on the software controlling the ICS (as we saw, several control strategies are possible for ICSs such as the TEP). However, designing security-aware control strategies that also meet operational and economic requirements is cumbersome~\cite{DBLP:conf/nordsec/KrotofilC13} and completely avoiding certain weakness patterns could even be impossible in certain settings.

If all of the above fails and the attacker is able to perform the identified near-optimal single-shot attack, then an operator can try to detect the attack as soon as possible by using monitoring systems to detect manipulated sensor values~\cite{DBLP:conf/ccs/CardenasALHHS11}. However, the detection time of such systems can be in the order of hours, which would be too slow to detect near-optimal attacks which, depending on the concrete case, can potentially bring down the ICS within minutes. An idea to improve the detection would be to tighten the monitoring specifically at those components in the ICS that our CCL-graph based attack method matches with known software weaknesses.

\section{Related Work}
Several researchers have analyzed different security aspects of the TEP.
Here we focus in a subset of those works that have studied diverse attacks against the TEP.
In~\cite{DBLP:conf/nordsec/KrotofilC13}, the goal was to get insights on the resilience of the physical process under attack.
Similar to our attacker model, they focused on compromising sensors and identifying those in best capacity to cause a shutdown on the targeted infrastructure.
In this particular case, the targeted infrastructure was Larsson's implementation of the TEP, described in our evaluation as control strategy \#1.
Aligned with our results, they found out that sensors 4, 9, and 17 are the best targets under the minimum and maximum value attack strategies.
However, they reached that conclusion by using much more knowledge than our approach, i.e., a fully-fledged simulation of the targeted infrastructure.
For an attacker to be able to build an accurate simulation of the targeted infrastructure, he would need access to at least, full PLC(s) configuration, P\&IDs, and documentation about the system's constraints.

In another work, the goal was to find out the right time to launch an attack on individual sensor signals to cause a shutdown of the targeted infrastructure~\cite{DBLP:conf/acsac/KrotofilCML14}.
In this case, again, the target was Larsson's implementation of the TEP.
The authors focused on DoS attacks on sensor signals, which forces control functions to use a stale value from the sensor. 
Under the assumption that launching sensor attacks at minimum or maximum peaks is the fastest way to cause a shutdown of the plant, the goal was to identify such peaks in real time.
They approached this challenge using the Best Choice Problem (BCP) methodological framework.
Using different learning windows, they identified sensors 4, 9, and 17 as the best candidate targets (i.e., fastest SDT).
To execute this approach a potential attacker might need at least network traffic access, a notion about the physical process and its potential disturbances (e.g., from P\&IDs), and ideally, some historic data.
Moreover, this approach has three main disadvantages besides the increased attacker knowledge about the plant.
First, it is focused on identifying the right time to launch an attack but does not indicate the optimal target. Launching an attack on the wrong target might not cause any impact whatsoever on the plant.
Only after executing attacks on all sensors it becomes clear which of them are optimal, which is unrealistic.
Second, the time to attack is lengthy (this includes the learning window and the selection of the moment to launch the attack), ranging from 9.61 hours to 27.17 hours in their experiments.
Third, in some circumstances their approach might not reach a conclusion, in which case the attacker ``has to choose a clearly suboptimal candidate (last sample in the attack window) or decide to not launch an attack.''~\cite{DBLP:conf/acsac/KrotofilCML14}.

The works in~\cite{DBLP:conf/ccs/CardenasALHHS11} and~\cite{DBLP:journals/ijcip/HuangCALTS09}, analyze diverse attacks against a simplified version of the TEP.
Both works incorporate detailed knowledge from the process dynamics to execute optimal attacks.
An attacker leveraging the techniques proposed in these 2 works would require at least access to the
PLC(s) configuration, historic data, and documentation about the system's constraints.
We deem the simplified version of the TEP so small that we use it only in our motivating example (Sect.~\ref{subsec:motivating_example}).
However, as we showed, our approach is also applicable to this plant and our results match the results obtained by previous works, but using limited knowledge.

No previous security works have used the TEP implementation by Luyben et al.~\cite{luyben1999plantwide},
possibly because the code was only available in the Fortran programming language.
We hope that our contribution in translating the code to MATLAB eases the challenging task of evaluating ICS security research for future works.

\section{Conclusion}
In this work, we investigated if constrained attackers without detailed system knowledge and simulators can identify near-optimal attacks. In contrast to attacks in prior work (that require precise algorithms, operating parameters, process models or simulators), in our approach the attacker only requires abstract knowledge on general information flow in the plant. Based on that information, we construct a CCL graph, and apply graph-based pattern matching based on several weakness patterns from the CWE database. 

Our resulting approach provides us with one (or more) sensors to attack. Experimentally, we applied and validated our approach on two use cases, and demonstrated that the approach successfully generates single-shot attacks, i.e., near-optimal attacks that are reliably shutting down a system on the first try. 

Our positive results in finding near-optimal targets with limited knowledge suggest that the difficulty of preparing ICS attacks is lower than previously thought.
We not only showed that the graph analysis can be automated, but that the graph creation can be automated too (e.g., using BACnet network traffic).
This significantly lowers the bar for ICS attacks and calls for a reassessment of the actual security risks in ICSs.\medskip

\noindent \textbf{Acknowledgments.} This work is partially funded by the Costa Rica Institute of Technology.

\vfill

\pagebreak

\begin{subappendices}
\renewcommand{\thesection}{\Alph{section}}%
\section{Pattern Matching Queries in CQL}
\label{appendix:cql}
Cypher Query Language (CQL) queries used to implement the proposed target identification approach.

\lstset{language=SQL, frame=single, caption=Pre-processing stage: setting the in- and out-degree centrality to variable nodes., captionpos=b, label=lst:degree}
\begin{lstlisting}
MATCH (n)
WHERE NOT n:FUNCTION_TYPE
SET n.indegree = size((n)<-[]-());

MATCH (n)
WHERE NOT n:FUNCTION_TYPE
SET n.outdegree = size((n)-[]->());
\end{lstlisting}

\lstset{language=SQL, frame=single, caption=Global variable pattern matching., captionpos=b, label=lst:global}
\begin{lstlisting}
MATCH (n:NODE_TYPE)
WITH AVG(n.outdegree) as average,
     stDev(n.outdegree) as stdev
  MATCH(m:NODE_TYPE)
  WHERE m.outdegree > (average+stdev)
  RETURN m;
\end{lstlisting}

\lstset{language=SQL, frame=single, caption=Multi-purpose variable pattern matching., captionpos=b, label=lst:multi}
\begin{lstlisting}
MATCH(n:NODE_TYPE)
WHERE n.indegree >= 2
RETURN n;
\end{lstlisting}

\lstset{language=SQL, frame=single, caption=Finding paths from sensors to non-sensor nodes chosen during the pattern matching phase., captionpos=b, label=lst:path}
\begin{lstlisting}
MATCH (src:Sensor), (dest {id:NODE_ID})
WHERE (src)-[*]->(dest)
RETURN src;
\end{lstlisting}

\pagebreak

\section{CCL Graph of Control Strategy \#1}
\label{appendix:control_strategy_1}
\begin{figure}[H]
  \centering
  \includegraphics[width=\textwidth]{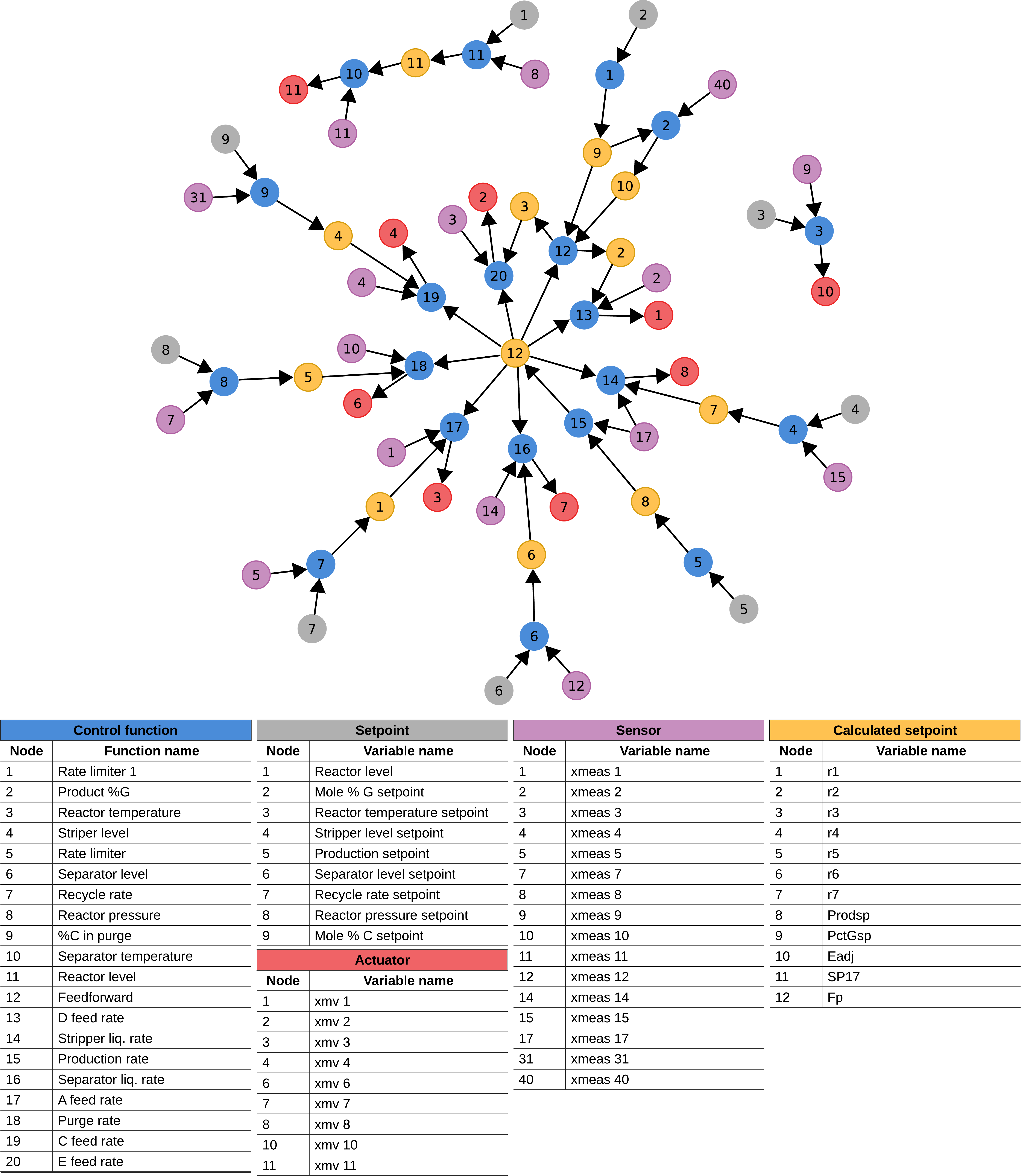}
  \caption{CCL graph of control strategy \#1. The color of each node represents its type, as described in Figure~\ref{fig:ccl_combinations}. The name of each node corresponds to the variable name in the original MATLAB code.}
  \label{fig:Larsson_graph}
\end{figure}

Figure~\ref{fig:Larsson_graph} shows the CCL graph of the first control strategy, proposed by Larsson et al.~\cite{larsson2001self}. This control strategy uses 16 sensors and 9 actuators from those available in the plant.
It also uses 9 static setpoints, 20 control functions, and 12 calculated setpoints. The resulting CCL graph is comprised of 66 nodes.

\section{CCL Graph of Control Strategy \#2}
\label{appendix:control_strategy_2}
\begin{figure}[H]
  \centering
  \includegraphics[width=\textwidth]{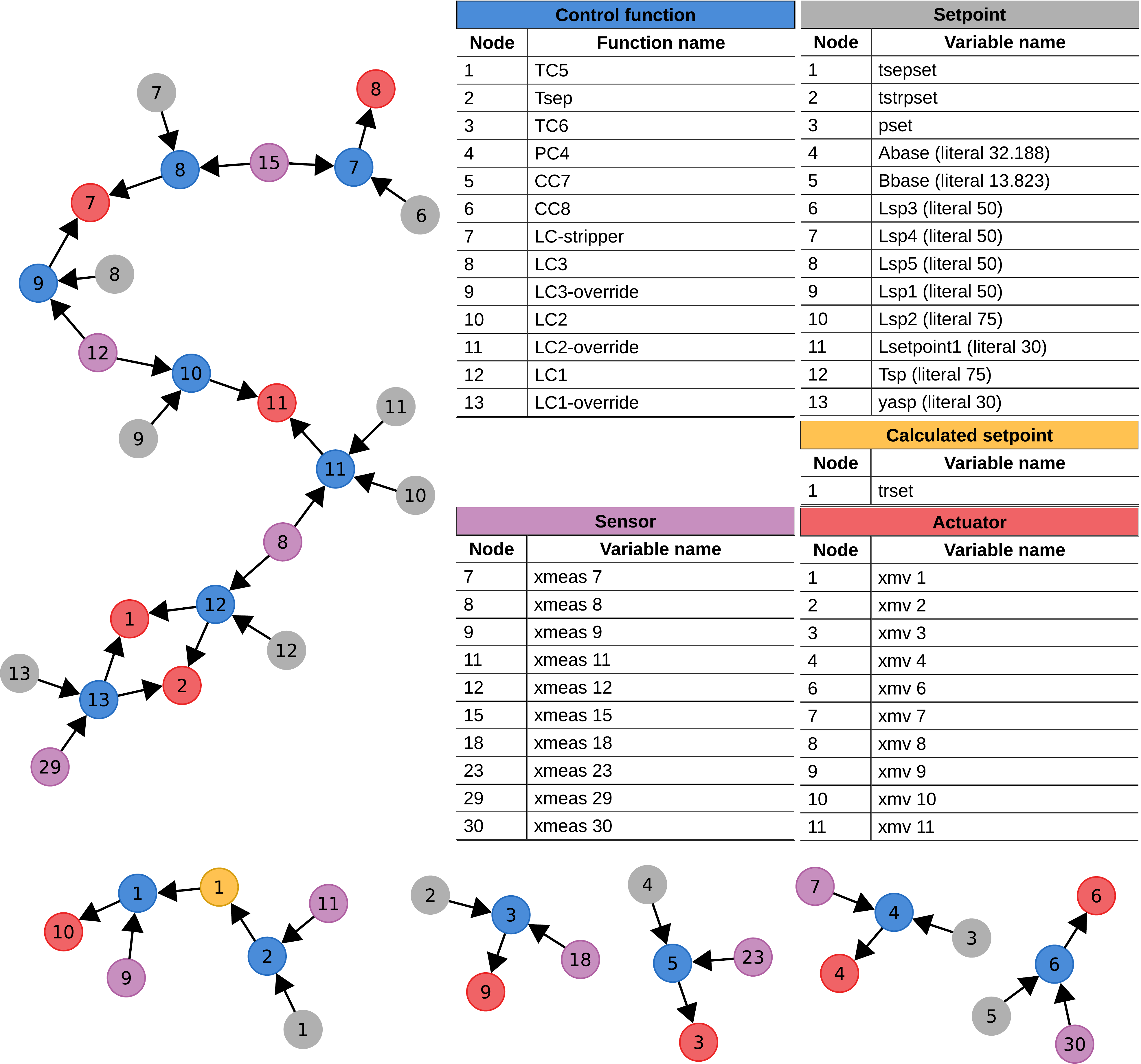}
  \caption{CCL graph of control strategy \#2. The color of each node represents its type, as described in Figure~\ref{fig:ccl_combinations}. The name of each node corresponds to the variable name in the original Fortran code.}
  \label{fig:Luyben_graph}
\end{figure}

Figure~\ref{fig:Luyben_graph} shows the CCL graph of the second control strategy, proposed by Luyben et al.~\cite{luyben1999plantwide}. This control strategy uses 10 sensors and 10 actuators from those available in the plant.
It also uses 13 static setpoints, 13 control functions, and 1 calculated setpoint. The resulting CCL graph is comprised of 47 nodes.

\pagebreak

\section{BACnet System}
\label{appendix:BACnet}
\begin{figure}[H]
  \centering
  \subfloat[\label{fig:global_setpoint}A static setpoint used by 14 control functions.]
           {\includegraphics[width=2.2in]{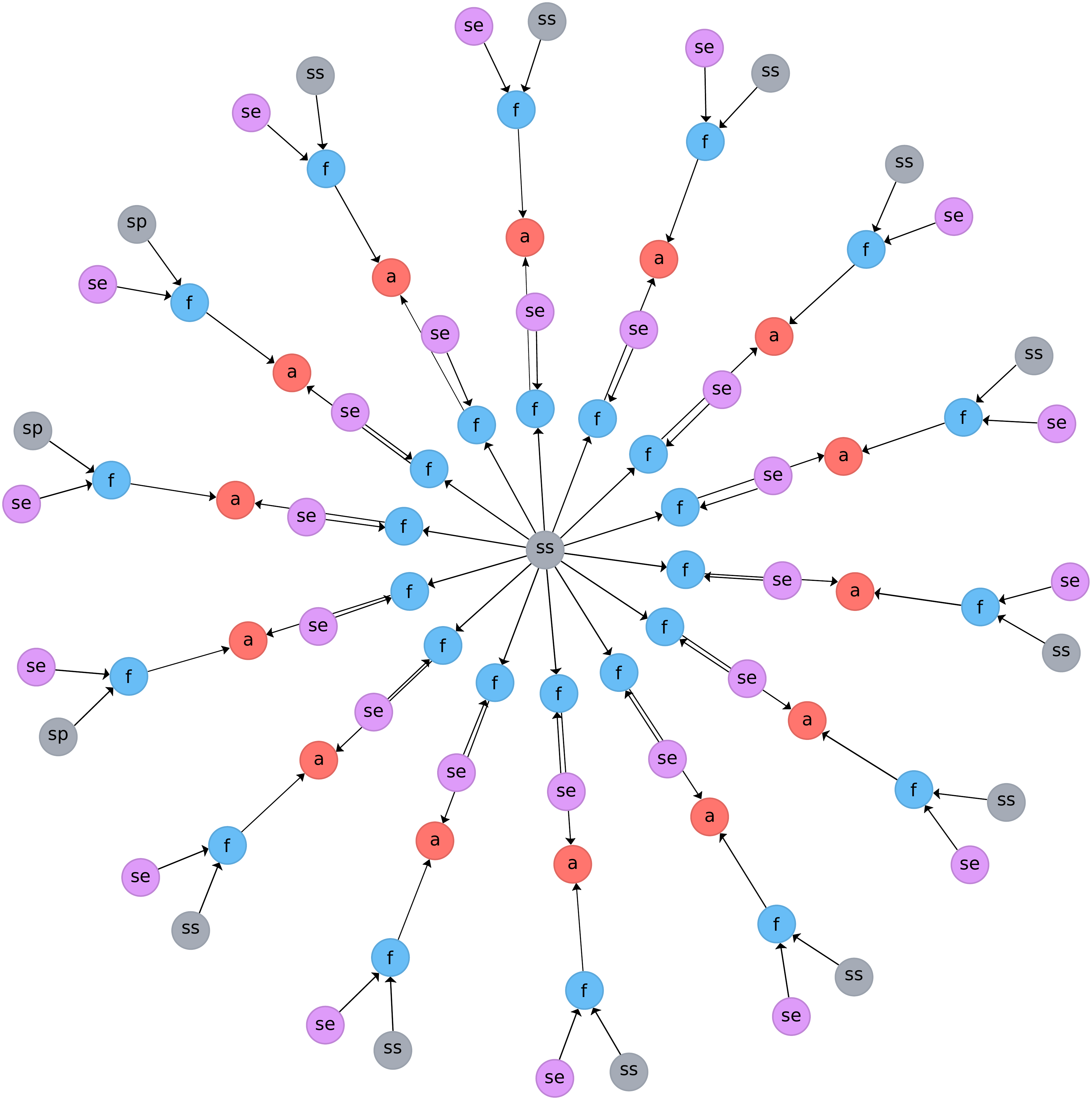}}\\
  \subfloat[\label{fig:act_set}Complex structure resembling a circular dependency.]
           {\includegraphics[width=2.0in]{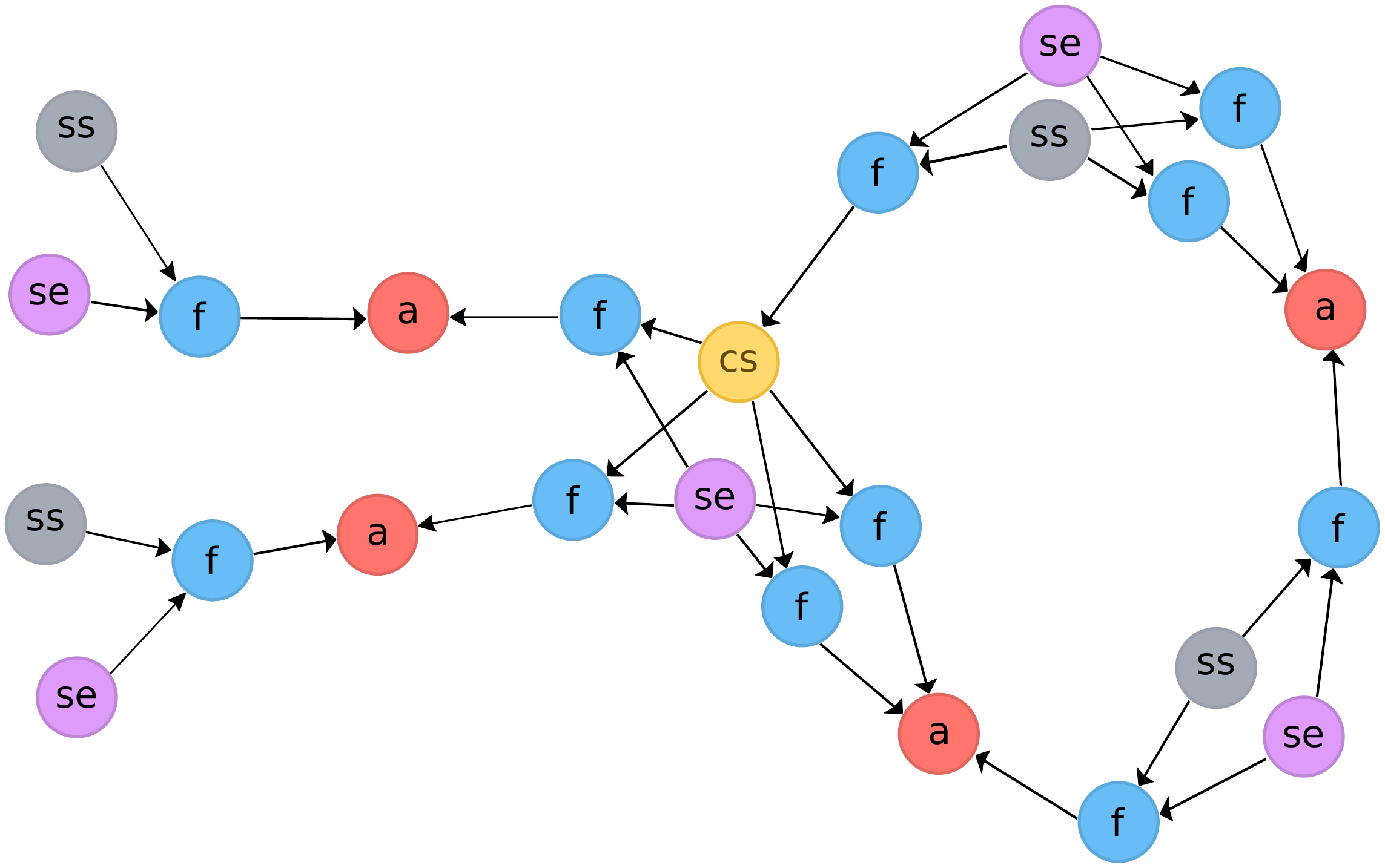}}\\
  \subfloat[\label{fig:actuator}An actuator commanded by 4 control functions.]
           {\includegraphics[width=2.0in]{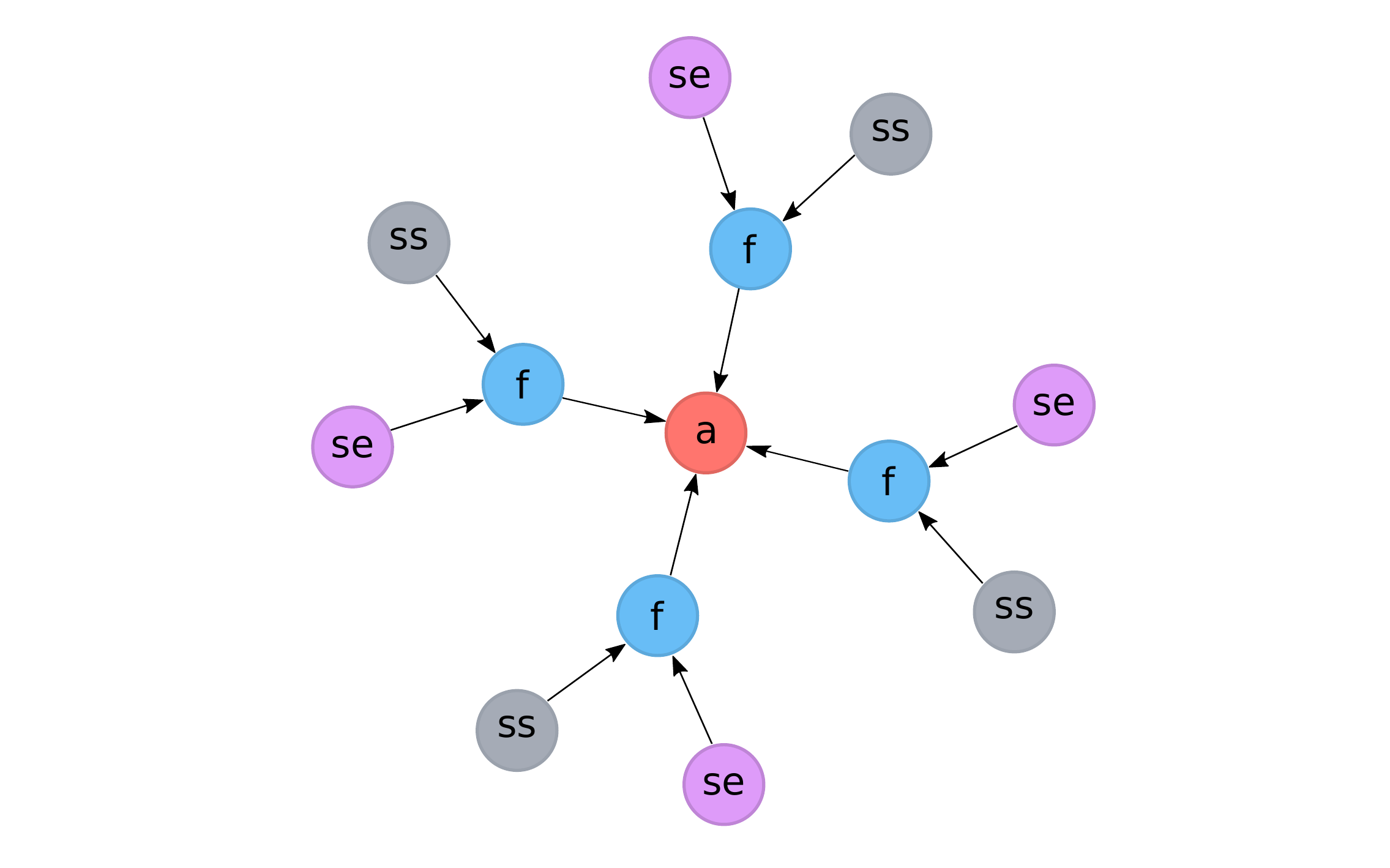}}\\
           \caption{Three subgraphs automatically extracted from a real and operational BACnet system.}
  \label{fig:real_bas}
\end{figure}

\begin{figure}[H]
  \centering
  \includegraphics[width=3.0in]{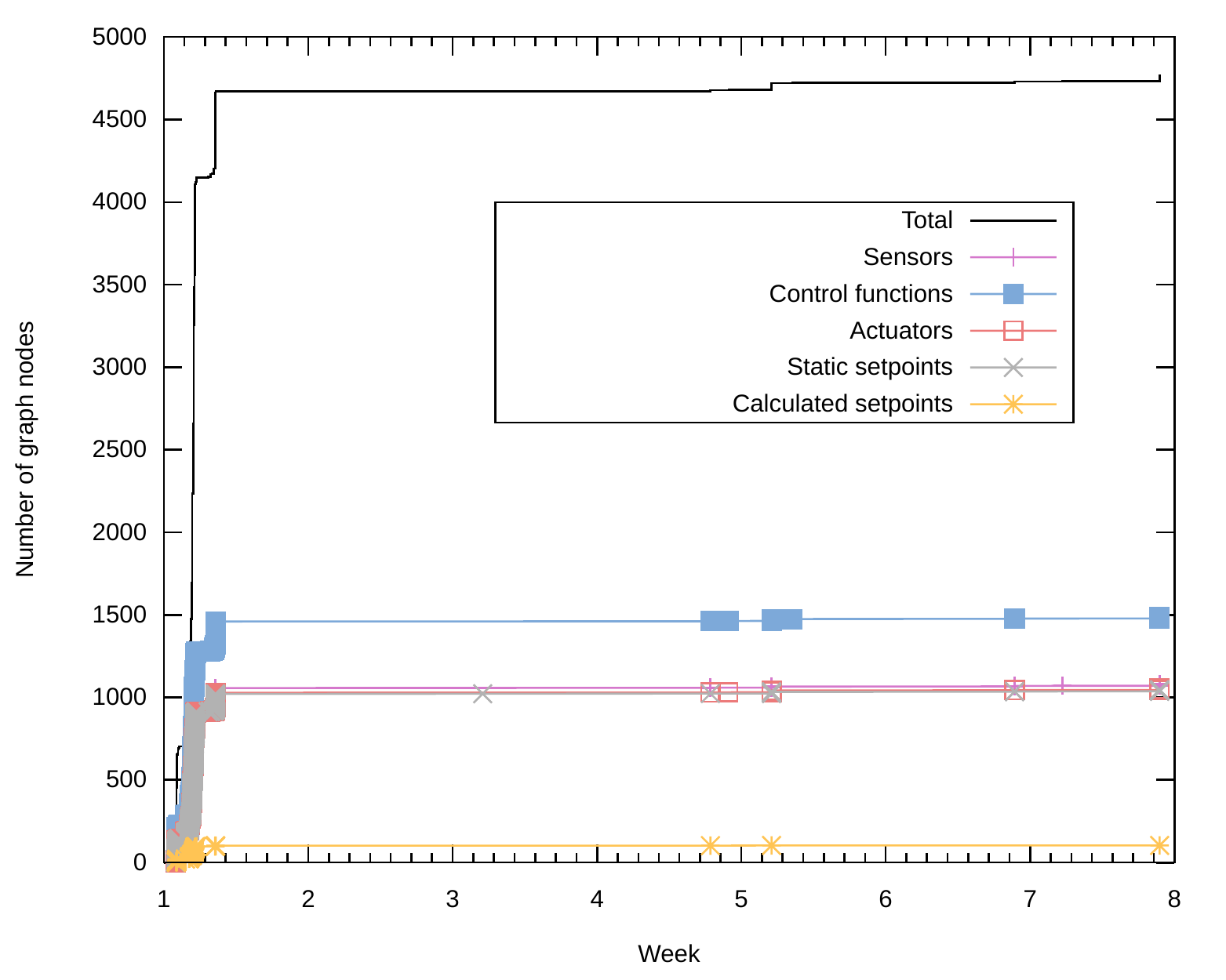}
  \caption{Node discovery rate during a 7-week long analysis of real BACnet traffic.}
  \label{fig:discovery_rate}
\end{figure}

In Figure~\ref{fig:real_bas}, we visualize three subgraphs we automatically extracted from real-world BACnet traffic.
In Figure~\ref{fig:discovery_rate}, we provide a plot of the node discovery rate during our experiment.

\section{Tennessee Eastman Plant}
\label{appendix:tep_diagram}
\begin{figure}[H]
  \centering
  \includegraphics[width=3.5in]{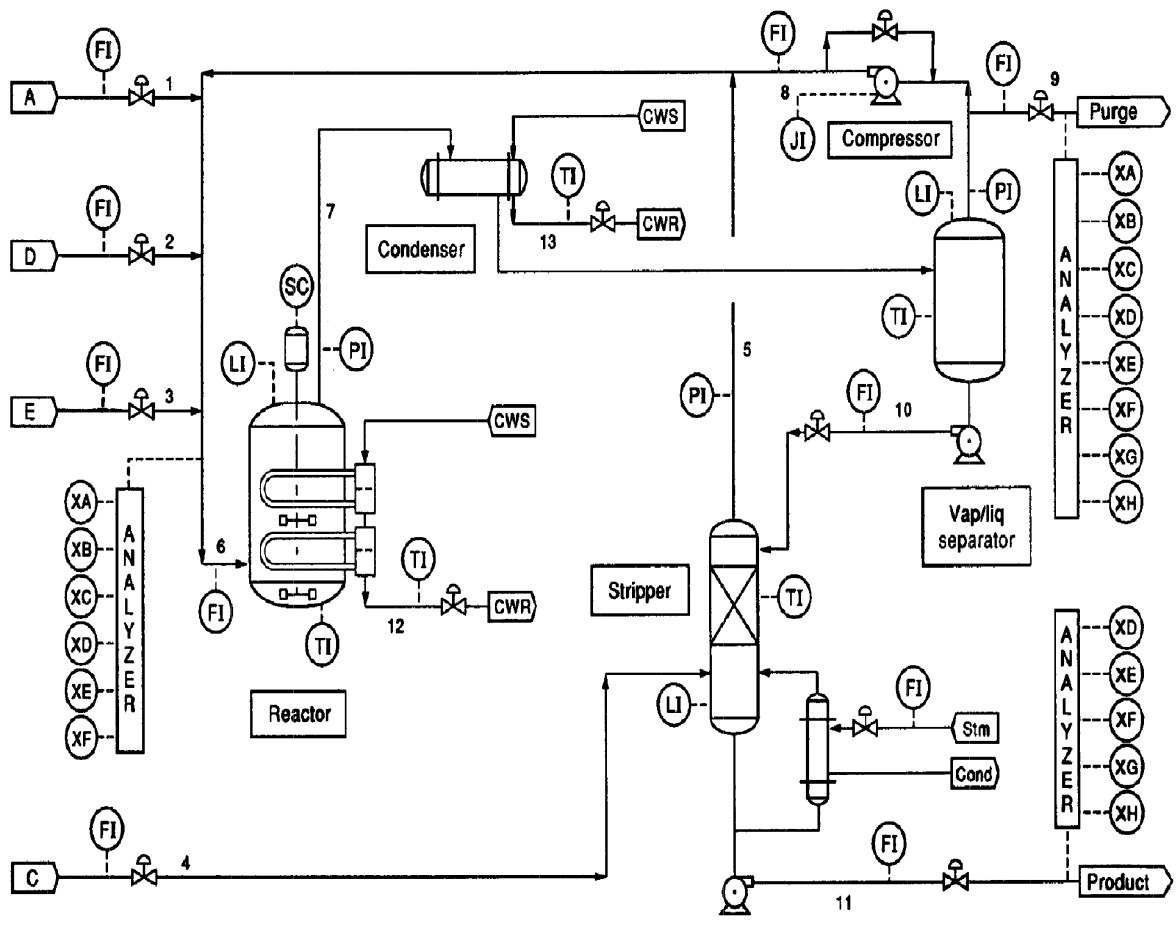}
  \caption{Tennessee Eastman Plant diagram taken from the original
    publication~\cite{downs1993plant}.}
  \label{fig:tep}
\end{figure}

Figure~\ref{fig:tep} shows a diagram of the hardware components available in the Tennessee Eastman Plant.

\end{subappendices}

\bibliographystyle{splncs04}
\bibliography{paper-bibliography}

\end{document}